\documentclass[showpacs,preprintnumbers,amsmath,amssymb,showkeys]{revtex4}

\usepackage[dvips]{graphicx}
\usepackage{dcolumn}
\usepackage{bm}

\newcommand{\nn}{\nonumber}

\def\Kv{\mbox{\boldmath $K$}}

\def\Pv{\mbox{\boldmath $P$}}

\def\qv{\mbox{\boldmath $q$}}
\def\kv{\mbox{\boldmath $k$}}

\begin{document}

\preprint{HUPD-1001, KUNS-2283, UTHEP-611}

\title{Resummation of large logarithms in the heavy quark
       effects \\ on the parton distributions inside the virtual photon}

\author{Yoshio Kitadono}
 \email{kitadono@theo.phys.sci.hiroshima-u.ac.jp}
 \affiliation{ Department of Physical Science, Faculty of Science,
               Hiroshima University,\\
               Higashi Hiroshima 739-8526, Japan.}

\author{Ryo Sahara}
 \email{sahara@scphys.kyoto-u.ac.jp}
 \affiliation{%
  Department of Physics, Graduate School of Science, Kyoto University, \\
  Yoshida, Kyoto 606-8501, Japan.}

\author{Tsuneo Uematsu}
 \email{uematsu@scphys.kyoto-u.ac.jp}
 \affiliation{%
  Department of Physics, Graduate School of Science, Kyoto University, \\
  Yoshida, Kyoto 606-8501, Japan.}

\author{Takahiro Ueda}
 \email{tueda@hep.ph.tsukuba.ac.jp}
 \affiliation{%
  Graduate School of Pure and Applied Sciences, University of Tsukuba, \\
  Tsukuba, Ibaraki 305-8571, Japan.}

\date{\today}

\begin{abstract}
We discuss the resummation of the large logarithmic terms
appearing in the heavy quark effects on parton distribution functions inside
the virtual photon. We incorporate heavy quark mass effects by changing the
initial condition of the leading-order DGLAP evolution equation. In a certain
kinematical limit, we recover the logarithmic terms of the next-to-leading
order heavy quark effects obtained in the previous work.
This method enables us to resum the large logarithmic terms due to heavy
quark mass effects on the parton distributions in the virtual photon.
We numerically calculate parton distributions using the formulae
derived in this work, and discuss the property of the resummed heavy quark
effects.
\end{abstract}



\maketitle

\section{Introduction}
The Large Hadron Collider (LHC)~\cite{LHC} has restarted at the CERN for
the purpose of discovering the Higgs boson, the new physics beyond the
standard model, and investigating the detailed information for
the quark gluon plasma, the B meson decays and so on.
The precise measurement will
be needed to confirm the discovery of the Higgs boson and searching the
beyond standard model at the electron-positron collider like
International Linear Collider (ILC)~\cite{ILC} and Super KEK-B~\cite{KEKB}.
In such a case, we have to know the behaviour of quantum
chromodynamics (QCD) at high energies because of the largeness of QCD
corrections.

\begin{figure}
  \begin{center}
    \includegraphics[scale=0.32]{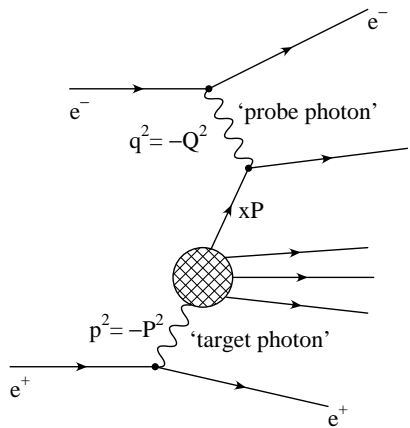}
    \caption{Deep inelastic scattering for virtual photons in $e^{+}e^{-}$
             collision.}
    \label{fig-thophotons}
 \end{center}
\end{figure}

There is a well-known fact that the cross section of the two-photon
processes
$e^{+}e^{-}\to \gamma^{*} \gamma^{*} \to e^{+}e^{-}+\mbox{hadrons}$
dominates over that of the one-photon annihilation processes
$e^{+}e^{-}\to \gamma^{*} \to \mbox{hadrons}$ in
the electron-positron collisions at high energies~\cite{twophoton}.
Let us consider the two-photon processes where both of the outgoing
$e^{+}$ and $e^{-}$ are detected and one of the virtual photons
is far off-shell with mass squared $q^2=-Q^2$, while the other photon
is close to the on-shell with mass squared $p^2=-P^2$.
In this kinematical region, the former photon is called
the `probe photon' and the latter one is
called the `target photon' (see Fig.~\ref{fig-thophotons}).

We can regard this two-photon process as
the deep-inelastic scattering in the electron-positron collision where
the target is a photon rather than a nucleon.
In this point of view,
we can define the photon structure functions as the analogues of the nucleon
structure functions, and the photon structure functions are predicted
by the quantum electrodynamics (QED) and QCD.
One of the difference between the nucleon structure functions and the
photon structure functions is the target mass ($P^2$) dependence.
The $P^2$ is not fixed for the virtual photon case, on the other hand,
$P^2$ is fixed for the nucleon case. There are many good reviews for the
theoretical as well as the experimental works for photon structure
functions, for example, see~\cite{Review}.

The real ($P^2=0$) unpolarised photon structure functions
$F^{\gamma}_{2}(x,Q^2)$ and $F^{\gamma}_{L}(x,Q^2)$
were investigated in the parton model (PM)~\cite{QPM},
in the perturbative QCD (pQCD) based on
the operator product
expansion (OPE)~\cite{CHM} supplemented by the renormalisation group (RG)
method~\cite{Witten,BB}, and also on the QCD improved PM \cite{Altarelli} powered by the
parton evolution equation~\cite{Dewitt,GR1983,MVV2002,MVV2004NNLOpart3}.
The real polarised photon structure function $g^{\gamma}_{1}(x,Q^2)$ was
investigated with pQCD in~\cite{polg1LO} for the leading-order (LO),
and for the next-to-leading order (NLO) in~\cite{polg1NLO1,GRS2001}.

The virtual ($P^2\neq 0$) unpolarised photon structure functions
$F^{\gamma}_{2}(x,Q^2,P^2)$ and $F^{\gamma}_{L}(x,Q^2,P^2)$
were also investigated by~\cite{UW1,UW2,Rossi,BA,Chyla}
in the kinematical region
\begin{eqnarray}
 \Lambda^2 \ll P^2 \ll Q^2, \label{disregion}
\end{eqnarray}
where $\Lambda$ is the fundamental QCD scale parameter.
The advantage to study the virtual photon target for the kinematical
region (\ref{disregion}) is that we can calculate the whole shape
and magnitude of the photon structure functions entirely by the
perturbative method.
Based on the recent results for the three-loop calculation for
the photon-quark and the photon-gluon splitting functions
(anomalous dimensions)~%
\cite{MVV2004NNLOpart3,MVV2004NNLOpart1,MVV2004NNLOpart2},
the unpolarised virtual photon structure function
$F^{\gamma}_{2}(x,Q^2,P^2)$ ($F^{\gamma}_{L}(x,Q^2,P^2)$)
was studied to the NNLO (to the NLO)~\cite{USU2007,KSUU2008},
and the polarised virtual photon structure function
$g^{\gamma}_{1}(x,Q^2,P^2)$ was studied to the NLO in~%
\cite{GRS2001,SU1999,BSU2002,SUU2006}.

In the parton picture, the photon structure function is expressed as
the convolution of the parton distribution function (PDF) in the
virtual photon with the coefficient functions in the OPE formalism.
We can also give the definite prediction for the PDFs inside the photon,
which we call \lq\lq photon PDFs\rq\rq \ for short, in
the present paper.
The theoretical calculations were done for unpolarised and
polarised photon PDFs in~%
\cite{Rossi,DG,GRStratmann,Fontannaz,SU2000,USU2009}.
However, these calculations in~\cite{MVV2004NNLOpart3,USU2007,SU2000,USU2009}
were assumed that all quarks in the virtual
photon are massless.
When the centre of mass energy is enough large to produce
heavy quarks (with mass $m$), i.e. $(p+q)^2 \ge 4m^2$,
the heavy quark mass effects should be taken into account. 
In the case of the nucleon target, the heavy quark mass effects were studied
by a method based on the OPE in \cite{Buza-etal1999}.

Many authors have investigated the heavy quark mass effects
in the photon structure functions~\cite{GR1983,GRS2001,GRStratmann,Fontannaz,AFG,GRSch1999,SSU2002,CJKL2003,CJK2004,GRV1992b,KSUU2009}.
The heavy quark mass effects were included in
the theoretical calculation for the PDFs of the real photon in Ref.~%
\cite{Fontannaz,AFG} by changing the initial condition of the DGLAP equation,
for the virtual photon PDFs in Ref.~\cite{KSUU2009,KSUU2010}
by using the OPE formalism supplemented by the mass-independent DGLAP equation.
In Refs.~\cite{Fontannaz,AFG}, the heavy quark mass effects are incorporated
by changing the initial condition. On the other hand,
in Refs.~\cite{KSUU2009,KSUU2010}, the heavy quark mass effects are included by
evaluating the finite matrix elements of the heavy-quark operators between the
photon states.
Recently we have found that the DGLAP equation with modified initial
condition for the heavy quark PDF~\cite{Fontannaz,AFG}
leads to the similar results which we have obtained by the OPE method~%
\cite{KSUU2009,KSUU2010} as we have mentioned before.

In the present paper, we adopt the alternative way to treat the heavy quark
mass effects on the virtual photon PDFs by setting the initial condition
for the heavy quark PDF in the DGLAP evolution equation,
which amounts to sum up the large logarithmic terms.
Through this prescription we try to improve the previous results for the heavy
quark mass effects.

In the next section, we discuss the basic formalism of the
evolution equation for virtual photon PDFs. We derive the explicit
expressions for the virtual photon PDFs with the heavy quark mass
effects in which the large logarithmic terms are resummed in
section \ref{modification}. In section \ref{numerical},
we present the numerical calculation of the results for photon PDFs.
The final section is devoted to the conclusions.

\section{Basic Formalism}
\label{basic}
First let us discuss the basic formalism of the evolution equation for
the PDFs in the virtual photon to the leading-order (LO) in QCD.
We consider the system with $n_f$ quarks and decompose it into two groups,
$n_f-1$ {\it light} (i.e. massless) quarks and a {\it heavy} quark.
Let
\begin{eqnarray}
 q^i_L(x,Q^2,P^2), \qquad q^\gamma_H(x,Q^2,P^2), \qquad
G^\gamma(x,Q^2,P^2), \qquad \Gamma^{\gamma}(x,Q^2,P^2) ,\label{photonicPDF}
\end{eqnarray}
be light quark distributions (with $i$ flavour and $i=1,\ldots,n_f-1$),
heavy quark distribution, gluon distribution and photon distribution
inside the virtual photon, respectively.
All the PDFs $q_{i}^{\gamma}(x, Q^2, P^2)$ evolve from the
virtuality of the target photon $P^2$ to the virtuality of the probe photon $Q^2$.
At the LO of the QED coupling constant
($\mathcal{O}$($\alpha$); $\alpha=e^2/4\pi$),
$\Gamma^{\gamma}(x,Q^2,P^2)$ does not evolve with respect to $Q^2$
and therefore we set $\Gamma^{\gamma}(x,Q^2,P^2)=\delta(1-x)$.
Since light quarks are distinguished from other quarks only through their
electromagnetic charge,
it is good to change the flavour basis to the light singlet and the
light nonsinglet.
We define the light singlet $q^{\gamma}_{Ls}$ and the light nonsinglet
$q^{\gamma}_{Lns}$ by the equations
\begin{eqnarray}
 q^{\gamma}_{Ls}
  \equiv \sum_{i=1}^{n_f - 1} q^{i}_L~,
\qquad q^\gamma_{Lns}
\equiv
 \sum_{i=1}^{n_f - 1} e^2_i
  \Bigl(q^{i}_L - \frac{1}{n_f - 1}q^{\gamma}_{Ls}\Bigr) , \label{qLs&qiLns}
\end{eqnarray}
where $e_{i}$ is the electromagnetic charge for $i$-th flavour quark
in the unit of proton charge.
The photon PDFs are described by a row vector $\qv^{\gamma}$ which
satisfies the inhomogeneous DGLAP evolution equation
\begin{eqnarray}
 \frac{d}{d\ln Q^2} \qv^{\gamma}(x, Q^2, P^2)
 &=& \int_{x}^{1} \frac{dy}{y} \qv(y,Q^2,P^2)
     \hat{\Pv}\left(\frac{x}{y}, Q^2\right)
 +  \kv(x,Q^2) , \label{dglap}
\end{eqnarray}
where the row vector $\qv^{\gamma}(x, Q^2, P^2)$ is
defined as
\begin{eqnarray}
 \qv^{\gamma}
= \left(
 q^{\gamma}_{Ls}, q^\gamma_H, G^\gamma, q^{\gamma}_{Lns}
\right),
\end{eqnarray}
and another row vector $\kv^{\gamma}(x, Q^2)=( k_{Ls}, k_{H}, k_{G}, k_{Lns})$
denotes the photon-parton splitting functions.
The $4 \times 4$ matrix
$\hat{\Pv}(z,Q^2)$ is expressed as
\begin{eqnarray}
 \hat{\Pv}(z,Q^2)
 =
 \begin{pmatrix}
   P_{LL}^{S}(z,Q^2) & P_{HL}(z,Q^2) & P_{GL}(z,Q^2) & 0 \\
   P_{LH}(z,Q^2)     & P_{HH}(z,Q^2) & P_{GH}(z,Q^2) & 0 \\
   P_{LG}(z,Q^2)     & P_{HG}(z,Q^2) & P_{GG}(z,Q^2) & 0 \\
   0                 & 0             & 0             & P_{LL}^{NS}(z,Q^2)
 \end{pmatrix},
\end{eqnarray}
where each element $P_{AB}$ means a splitting function of $B$ parton to
$A$ parton.

One can solve the DGLAP equation (\ref{dglap}) by introducing the moments
of the photon PDFs~\cite{QPM}. Here we discuss the procedure briefly.
By taking the moment, one obtains the equation
\begin{eqnarray}
 \frac{d}{d\ln Q^2} \qv^{\gamma}(n, Q^2, P^2)
 &=& \kv(n,Q^2) + \qv^{\gamma}(n,Q^2,P^2)
     \hat{\Pv}\left(n, Q^2 \right),
  \label{dglap-n}
\end{eqnarray}
where an $n$-th moment $f(n)$ of a function $f(x)$ is defined by
\begin{eqnarray}
 f(n) \equiv \int_{0}^{1}dx x^{n-1} f(x).
\end{eqnarray}
Then we introduce the variable $t$ as~\cite{FP1}
\begin{eqnarray}
 t &=& \frac{2}{\beta_0} \ln \frac{\alpha_s(P^2)}
                                  {\alpha_s(Q^2)},
\end{eqnarray}
instead of $Q^2$.
We expand $\kv_{n}=\kv(n,Q^2)$,
$\hat{\Pv}_{n}=\hat{\Pv}(n,Q^2)$
and $\qv^{\gamma}_{n}=\qv^{\gamma}(n,Q^2,P^2)$
in powers of the QED coupling constant $\alpha$ as well as the QCD coupling
constant $\alpha_{s}$ as follows
\begin{eqnarray}
 \kv_{n} &=&   \frac{\alpha}{2\pi} \kv^{(0)}_{n}
             + \frac{\alpha}{2\pi} \frac{\alpha_{s}(t)}{2\pi}
	        \kv^{(1)}_{n}
             + \cdots, \\
 \hat{\Pv}_{n} &=&   \frac{\alpha_s(t)}{2\pi} \hat{\Pv}^{(0)}_{n}
             + \left( \frac{\alpha_{s}(t)}{2\pi} \right)^2
	        \hat{\Pv}^{(1)}_{n}
             + \cdots,\\
\qv^{\gamma}_{n} &=&   \qv^{\gamma (0) }_{n}
                     + \qv^{\gamma (1) }_{n} + \cdots,
\end{eqnarray}
where the $\alpha_s$ dependence appears in $\qv^{\gamma}_{n}$
implicitly with the notation~\cite{USU2007},
and $\qv^{\gamma (0)}_{n}$ and $\qv^{\gamma (1)}_{n}$ correspond to the LO
and NLO solutions, respectively.
One finally obtains the LO solution $\qv^{\gamma (0)}_{n}$ for the DGLAP
equation (\ref{dglap-n}) as
\begin{eqnarray}
 \qv^{\gamma (0)}_{n}(t)
 &=&  \frac{\alpha}{8\pi\beta_0} \frac{4\pi}{\alpha_{s}(t)}
      \Kv_{n}^{(0)} \sum_{i} P_{i}^{n} \frac{1}{1+d_{i}^{n}}
      \left\{ 1 - r^{1+d_{i}^{n}} \right\}
    + \qv^{\gamma (0)}_{n}(0)\sum_{i}r^{d_{i}^{n}} P_{i}^{n}, \label{LOsolution}
\end{eqnarray}
where
$r$ is the ratio of
QCD couplings which is defined by
\begin{eqnarray}
 r  = \frac{\alpha_{s}(t)}{\alpha_{s}(0)}
    = \frac{\alpha_{s}(Q^2)}{\alpha_{s}(P^2)}.
\end{eqnarray}

Now the last term of (\ref{LOsolution}) is determined by the initial condition.
Although we usually set $\qv^{\gamma (0)}_{n}(0)=0$, as we took in our previous
paper~\cite{KSUU2010},
we change the initial condition for the heavy quark PDF
and we will discuss the relation between this modification and resummation
in the next section.

$\beta_{0}$ and $\beta_{1}$ appears in the perturbative expansion
of the QCD running coupling constant
of the QCD running coupling constant
\begin{eqnarray}
 \frac{d}{d\ln Q^2} \alpha_{s}(Q^2)
 &=& - \beta_{0} \frac{\alpha_{s}^2(Q^2)}{4\pi}
     - \beta_{1} \frac{\alpha_{s}^3(Q^2)}{(4\pi)^2} + \cdots, \label{run}
\end{eqnarray}
with $\beta_{0}=11-2n_f/3$ and $\beta_{1}=102-38n_f/3$.
The relation between $ \kv_{n}^{(0)}$ and $ \Kv_{n}^{(0)}$ is given
by,
\begin{eqnarray}
 \kv_{n}^{(0)} &=& \frac{1}{4} \Kv_{n}^{(0)},
\end{eqnarray}
where the elements of row vector
$\Kv_{n}^{(0)}=\left(K_{Ls}^{0,n}, K_{H}^{0,n},
K_{G}^{0,n},K_{Lns}^{0,n}\right)$ are evaluated to be
\begin{eqnarray}
K_{Ls}^{0,n} &=& 24 (n_{f} - 1) \langle e^2 \rangle_{L} k_{n}^{0},\\
K_{H}^{0,n}  &=& 24 e_{H}^2 k_{n}^{0},\\
K_{G}^{0,n} &=& 0,\\
K_{Lns}^{0,n} &=& 24 (n_{f} - 1)
 (\langle e^4 \rangle_{L} - \langle e^2 \rangle_{L}^2 ) k_{n}^{0},\\
k_{n}^{0} &=& \frac{n^2+n+2}{n(n+1)(n+2)},
\end{eqnarray}
where the charge factors for quadratic term $\langle e^2 \rangle_{L}$ and for
quartic term $\langle e^4 \rangle_{L}$ are defined by
\begin{eqnarray}
   \langle e^2 \rangle_{L} &\equiv& \frac{1}{n_f-1}
  \sum_{i=1}^{n_f-1} e_{i}^{2},
\hspace{2cm}
  \langle e^4 \rangle_{L} \equiv \frac{1}{n_f-1}
  \sum_{i=1}^{n_f-1} e_{i}^{4}.
\end{eqnarray}
Note that $K_{Ls}^{0,n}+K_{H}^{0,n}=K_{\psi}^{0,n}$,
and $K_{\psi}^{0,n}$ is the usual flavour-singlet anomalous dimension
for $n_f$ massless quarks.
The relation between $\hat{\Pv}_{n}^{(0)}$ and $P_{i}^{n}$ is given by
\begin{eqnarray}
  \hat{\Pv}_{n}^{(0)}
 &=& - \frac{1}{4} \hat{\gamma}^{(0)}_{n}
  =  - \frac{1}{4} \sum_{i=\psi,\pm,Lns}\lambda_{i}^{n} P_{i}^{n},
\end{eqnarray}
where $\hat{\gamma}^{(0)}_{n}$ is the one-loop hadronic anomalous dimension
matrix, $\lambda_{i}^{n}=2\beta_{0}d_{i}^{n}$ ($i=\psi,\pm,Lns$) are
the eigenvalues of $\hat{\gamma}^{(0)}_{n}$
which are given by
\begin{eqnarray}
 \lambda_{\psi}^n &=& \lambda_{Lns}^n = \gamma_{\psi\psi}^{0,n},\\
 \lambda_{\pm}^n &=&
  \frac{1}{2}
  \left[ \gamma_{\psi\psi}^{0,n} + \gamma_{GG}^{0,n}
         \pm \sqrt{ (\gamma_{\psi\psi}^{0,n} - \gamma_{GG}^{0,n})^2
	           + 4 \gamma_{\psi G}^{0,n} \gamma_{G\psi}^{0,n}}
	  \right].
\end{eqnarray}
The $P_{i}^{n}$ are the projection matrices in the spectral decomposition for
$\hat{\gamma}^{(0)}_{n}$ which satisfy the following relations:
\begin{eqnarray}
 P_{i}^n P_{j}^n &=& \delta_{ij} P_{i}^n,~\hspace{1cm}
 \sum_{i}P_{i}^{n} = \bf{1}_{4\times4},
\end{eqnarray}
where $i,j =\psi, \pm, Lns$ runs over the eigenvalues, and
the explicit expressions are given by
\begin{eqnarray}
 P_\psi^n &=&
\left(
\begin{array}{ccc|c}
\frac{1}{n_f}\ &-\frac{1}{n_f}\ &0&0\\
-\frac{n_f-1}{n_f}\ &\frac{n_f-1}{n_f}\ &0&0\\
0&0&0&0\\
\hline
0&0&0&0
\end{array}
\right)~,\hspace{2cm}
\hspace{-0.5cm} P_{Lns}^n =
\left(
\begin{array}{ccc|c}
0\  &0\ &0\ &0 \\
0\  &0\ &0\ &0 \\
0\  &0\ &0\ &0 \\
\hline
0&0&0&1
\end{array}
\right),~\\
\hspace{-0.5cm}P_\pm^n &=&
\frac{1}{\lambda_\pm^n-\lambda_\mp^n}
\left(
\begin{array}{ccc|c}
\frac{n_f-1}{n_f}(\gamma_{\psi\psi}^{0,n}-\lambda_\mp^n) &
\frac{1}{n_f}(\gamma_{\psi\psi}^{0,n}-\lambda_\mp^n)\ &\gamma_{G\psi}^{0,n}&0\\
\frac{n_f-1}{n_f}(\gamma_{\psi\psi}^{0,n}-\lambda_\mp^n) &
\frac{1}{n_f}(\gamma_{\psi\psi}^{0,n}-\lambda_\mp^n)\ &\gamma_{G\psi}^{0,n}&0\\
\frac{n_f-1}{n_f}\gamma_{\psi G}^{0,n}\ &\frac{1}{n_f}\gamma_{\psi G}^{0,n}
&\gamma_{GG}^{0,n}-\lambda_\mp^n&0\\
\hline
0&0&0&0
\end{array}
\right)~.
\end{eqnarray}
The most important part in the present
paper is $\qv_{n}^{\gamma(0)}(0)$, which is the
initial condition for the DGLAP Eq.~(\ref{dglap-n}).
We usually set this term $\qv^{\gamma (0)}_{n}(0)$ to be vanishing at the LO.
However, we will find that the resummed expression of the logarithm term
$\ln (m^2/P^2)$ which appears as the heavy quark mass effects in Eq.~(65)
in Ref.~\cite{KSUU2010} is recovered by changing the initial condition of the
DGLAP Eq.~(\ref{dglap-n}) for the heavy quark component as
discussed in the next section.

\section{Initial condition and resummation}
\label{modification}
In Ref.~\cite{KSUU2010} it is shown that the additional terms for the moment
of the PDF obtained by the OPE method
are proportional to the LO renormalisation group parameters and the
logarithmic terms.
Therefore we can expect the
possibility to derive the previous results by the LO evolution equation.
We demonstrate it by changing the initial condition for the LO evolution
equation.
\subsection{A change in the initial condition for the heavy quark}
The moment for the heavy quark component $q^{(0),n}_{H}(t)$
can be projected after some calculations by using the
Eq.~(\ref{LOsolution}),
\begin{eqnarray}
  q_{H}^{(0),n}(t) / \frac{\alpha}{8\pi\beta_0}
&=& \frac{4\pi}{\alpha_s(t)}
    \left[ - \frac{1}{1 + d_{\psi}^{n}}
             \left( \frac{1}{n_f} K_{\psi}^{n} - K_{H}^{0,n}
             \right)
	     \left( 1 - r^{d_{\psi}^{n} + 1} \right)
	   + \frac{1}{n_f} K_{\psi}^{0,n}
	     \sum_{\pm}\frac{1}{1 + d_{\pm}^{n}}
	     \frac{\gamma_{\psi\psi}^{0,n} - \lambda_{\mp}^{n}}
	          {\lambda_{\pm}^{n} - \lambda_{\mp}^{n}}
	     \left( 1 - r^{d_{\pm}^{n} + 1} \right)
    \right]\nn\\
&{}& \hspace{0.7cm} +
 \left[  \frac{n_f-1}{n_f} r^{d_{\psi}^{n}}
     + \frac{1}{n_f}
       \sum_{\pm}\frac{\gamma_{\psi\psi}^{0,n} - \lambda_{\mp}^{n}}
            {\lambda_{\pm}^{n} - \lambda_{\mp}^{n}}
       r^{d_{\pm}^{n}}
 \right]  \hat{q}_{H}^{(0),n}(0) / \frac{\alpha}{8\pi\beta_0},
\label{heavy-component}
\end{eqnarray}
where we denote $\sum_{\pm}f(\lambda_{\pm}) \equiv f(\lambda_+) + f(\lambda_-)$
and $\hat{q}_{H}^{(0),n}$ in the last term should be determined later.
As we mentioned previously, we change the initial condition of the DGLAP
equation.
Let us consider the following condition:
\begin{eqnarray}
 \qv^{\gamma (0)}_{n}(t=0)
&=& \left( 0, \hat{q}_{H}^{(0),n}, 0,  0
    \right),
\end{eqnarray}
where $\hat{q}_{H}^{(0),n}$ is the heavy-quark PDF evaluated
at the scale $t=0~(Q^2=P^2)$.
This modification for the initial condition of the LO DGLAP equation
corresponds to a kind of the heavy quark threshold effect, because the
evolution for the heavy quark component is suppressed by this condition.
The initial condition $\hat{q}_{H}^{(0),n}$ is determined by the equation
\begin{eqnarray}
 \hat{q}_{H}^{(0),n}(t_m) &=& 0,\\
 t_{m} &=& \frac{2}{\beta_0} \ln \frac{\alpha_s(P^2)}{\alpha_s(m^2)}.
\end{eqnarray}
Setting $t=t_{m}$ in Eq.~(\ref{heavy-component}), we obtain the following result:\begin{eqnarray}
 \hat{q}_{H}^{(0),n}(0) / \frac{\alpha}{8\pi\beta_0}
 &=& - H_{n}(t_m)/ J_{n}(t_m),
\end{eqnarray}
where the functions $H_{n}(t_m)$ and $J_{n}(t_m)$ are defined by
\begin{eqnarray}
 H_{n}(t_m)
&=& \frac{4\pi}{\alpha_s(t_m)}
    \left[ - \frac{1}{1 + d_{\psi}^{n}}
             \left( \frac{1}{n_f} K_{\psi}^{n} - K_{H}^{0,n}
             \right)
	     \left( 1 - r^{d_{\psi}^{n} + 1}_{m} \right)
\right. \nn\\
&{}& \left. \hspace{1.5cm}
	   + \sum_{\pm}\frac{1}{1 + d_{\pm}^{n}}
	     \frac{1}{n_f} K_{\psi}^{0,n}
	     \frac{\gamma_{\psi\psi}^{0,n} - \lambda_{\mp}^{n}}
	          {\lambda_{\pm}^{n} - \lambda_{\mp}^{n}}
	     \left( 1 - r^{d_{\pm}^{n} + 1}_{m} \right)
    \right],
\end{eqnarray}
\begin{eqnarray}
 J_{n}(t_{m})
 &=&   \frac{n_f-1}{n_f} r_{m}^{d_{\psi}^{n}}
     + \frac{1}{n_f} \sum_{\pm}
       \frac{\gamma_{\psi\psi}^{0,n} - \lambda_{\mp}^{n}}
            {\lambda_{\pm}^{n} - \lambda_{\mp}^{n}}
       r_{m}^{d_{\pm}^{n}}, \label{Jn}
\end{eqnarray}
and here $r_{m}=\alpha_s(t_m)/\alpha_s(0)$ corresponds to the ratio between
the QCD running coupling at the scale of the heavy quark mass
and that of the renormalisation scale for the photon matrix element in
OPE formalism. See the Appendix \ref{explicit} for the explicit expressions
of the PDFs in the virtual photon including the resummed mass effects.

If we denote the variation terms as $\Delta q_{i}^{\gamma(0),n}(t)$
which are due to the heavy quark effects, then we recover the fixed order
(NLO QCD + heavy quark mass effects) results as given by
\begin{eqnarray}
 \Delta q_{Ls}^{\gamma(0),n}(t) / \frac{\alpha}{8\pi\beta_0}
 &=&  \Delta \hat{A}_{Ls}^{n,\psi}
             \left( 1 - r^{ d_{\psi}^{n} } \right)
    + \sum_{\pm}\Delta \hat{A}_{Ls}^{n,\pm}
             \left( 1 - r^{ d_{\pm}^{n} } \right),\label{delta-qls}\\
 \Delta \hat{A}_{Ls}^{n,\psi}
 &=& \frac{n_f-1}{n_f} 2\beta_{0} \Delta \tilde{A}_{H}^{n}, \nn\\
 \Delta \hat{A}_{Ls}^{n,\pm}
 &=& - \frac{n_f-1}{n_f}
       \frac{\gamma_{\psi\psi}^{0,n} - \lambda_{\mp}^{n} }
            {\lambda_{\pm}^{n} - \lambda_{\mp}^{n} }
       2\beta_{0} \Delta \tilde{A}_{H}^{n}, \nn
\end{eqnarray}
\begin{eqnarray}
 \Delta q_{H}^{\gamma(0),n}(t) / \frac{\alpha}{8\pi\beta_0}
 &=&  \Delta \hat{A}_{H}^{n,\psi}
             \left( 1 - r^{ d_{\psi}^{n} } \right)
    + \sum_{\pm}\Delta \hat{A}_{H}^{n,\pm}
             \left( 1 - r^{ d_{\pm}^{n} } \right)
    + \Delta \hat{C}_{n},\label{delta-qh}\\
 \Delta \hat{A}_{H}^{n,\psi}
 &=& - \frac{n_f-1}{n_f} 2\beta_{0} \Delta \tilde{A}_{H}^{n}, \nn\\
 \Delta \hat{A}_{H}^{n,\pm}
 &=& - \frac{1}{n_f}
       \frac{\gamma_{\psi\psi}^{0,n} - \lambda_{\mp}^{n} }
            {\lambda_{\pm}^{n} - \lambda_{\mp}^{n} }
       2\beta_{0} \Delta \tilde{A}_{H}^{n}, \nn\\
\Delta \hat{C}_{n} &=& 2\beta_{0} \Delta \tilde{A}_{H}^{n}, \nn
\end{eqnarray}
\begin{eqnarray}
 \Delta G^{\gamma(0),n}(t) / \frac{\alpha}{8\pi\beta_0}
 &=&  \sum_{\pm}\Delta \hat{A}_{G}^{n,\pm}
             \left( 1 - r^{ d_{\pm}^{n} } \right),\\
 \Delta \hat{A}_{G}^{n,\pm}
 &=& - \frac{\gamma_{G\psi}^{0,n}}
            {\lambda_{\pm}^{n} - \lambda_{\mp}^{n} }
       2\beta_{0} \Delta \tilde{A}_{H}^{n}, \nn
\end{eqnarray}
\begin{eqnarray}
 \Delta q_{Lns}^{\gamma(0),n}(t) / \frac{\alpha}{8\pi\beta_0}
 &=&  0,\label{delta-qlns}
\end{eqnarray}
where $\hat{q}_{H}^{(0),n}(0) / \frac{\alpha}{8\pi\beta_0} = 2\beta_{0}
\Delta \tilde{A}_{H}^{n}$. The variation of operator matrix element
$\Delta \tilde{A}_{H}^{n}$ is evaluated from the LO coupling
$\alpha_s(t_m)$ together with the LO anomalous dimension $K_{H}^{0,n}$
in our method. Neglecting the finite term (large mass limit),
the variation of the operator matrix element is evaluated as
\begin{eqnarray}
 \Delta \tilde{A}_{H}^{n}
 &=& - \frac{1}{2} K_{H}^{0,n} \ln \frac{m^2}{P^2}
  =  - 12 e_{H}^2 \frac{n^2+n+2}{n(n+1)(n+2)}
       \ln \frac{m^2}{P^2}.
\end{eqnarray}
All the results in the large mass limit is consistent with that
of the variation terms in Ref.~\cite{KSUU2010} except for the mass-independent
finite terms. It is impossible to recover these terms only through the DGLAP
equation and these difference could be considered as the scheme-dependence
for photon PDFs.

\subsection{A certain limit}
Next we consider the expression by taking a limit
($\Lambda^2 \ll P^2 \ll m^2$) in order to compare our results with those
in Ref.~\cite{KSUU2010}.

In terms of the LO running coupling constant
\begin{eqnarray}
 \alpha_s(Q^2)
 = \frac{4\pi}{\beta_0} \frac{1}{\ln(Q^2/\Lambda^2)},
\end{eqnarray}
the ratio $r_{m}$ can be written as
\begin{eqnarray}
 r_{m} &=& \frac{\alpha_s(m^2)}{\alpha_s(P^2)}
        =  \frac{\ln(P^2/\Lambda^2)}{\ln(m^2/\Lambda^2)}
        = 1 - \frac{\ln(m^2/P^2)}{\ln(m^2/\Lambda^2)}
	\equiv 1 - \epsilon ,
\end{eqnarray}
where $\epsilon=\ln(m^2/P^2)/\ln(m^2/\Lambda^2)$.
Considering the case of large-mass limit:
$\ln(m^2/P^2) \ll \ln(m^2/ \Lambda^2$),
we can set the region of $\epsilon$ as $\epsilon \ll 1$.
Then we can expand $H_{n}(t_m), J_{n}(t_m)$ in the modified initial
condition term $q_{H}^{(0),n}$ up to $\mathcal{O}(\epsilon)$ as
\begin{eqnarray}
 H_{n}(t_m)
&=& \frac{4\pi}{\alpha_s(t_m)} K_{H}^{0,n}
\left[~\epsilon + O(\epsilon^2) ~\right],\\
 J_{n}(t_m)
 &=& 1 + O(\epsilon),
\end{eqnarray}
where we have used $(1+\epsilon)^{d} \approx 1 + d \epsilon$.
By using the fact
$\frac{4\pi}{\alpha_s(t_m)} = \beta_{0} \ln\frac{m^2}{\Lambda^2}$,
we obtain the result as
\begin{eqnarray}
  \hat{q}_{H}^{(0),n}(0) / \frac{\alpha}{8\pi\beta_0}
 \approx - \beta_{0} K_{H}^{0,n} \ln\frac{m^2}{P^2}
 \left[~1 + O(\epsilon)~\right],
\label{qh0expand}
\end{eqnarray}
where the order of the neglected term $O(\epsilon)$ corresponds to
the terms like $\ln^{k}(m^2/P^2), (k=1,2,3,\dots)$.
We recover the results about the large logarithmic term due to the
heavy quark mass effect which appears in the result by OPE formalism
except for mass-independent finite terms.
Therefore the equations (\ref{Ls}),(\ref{H}),(\ref{G}), and (\ref{Lns})
which we derived by the modification of the
initial condition for the heavy PDF in the LO DGLAP equation
are more general forms and the large logarithm terms are resummed.
The resummed terms form compact power terms $(r_{m})^{d_{i}^{n}}$ in $\hat{q}^{\gamma(0),n}_{H}(0)$.

\section{Numerical Results}
\label{numerical}
One can obtain the various photon PDFs from their moments
by the inverse Mellin transformation.
We show the results of numerical calculation for
the heavy quark PDF $q^{\gamma}_{H}$, the gluon PDF $G^{\gamma}$,
the light singlet quark PDF $q^{\gamma}_{Ls}$
and the effective photon structure function $F^{\gamma}_{\rm eff}$.
The last one, $F^{\gamma}_{\rm eff}$, is scheme-independent and is proportional to the total cross
section of the two photon process.
We consider the two cases which were measured in the experiments
~\cite{PLUTO,L3}.
The first case (i) resides in the PLUTO energy region
where we regard the charm quark as the heavy quark,
the second case (ii) is in the L3 energy region where we regard the
bottom quark as the heavy quark,
\begin{eqnarray}
\mbox{case (i)} &{}& \hspace{0.5cm}
        n_{f} = 4,
         ~Q^2 = 5~     \mbox{GeV}^2,
         ~P^2 = 0.35~  \mbox{GeV}^2,
       ~m_{c} = 1.3~   \mbox{GeV}, \\
\mbox{case (ii)} &{}& \hspace{0.5cm}
       n_{f} = 5,
        ~Q^2 = 120~   \mbox{GeV}^2,
        ~P^2 = 3.7~   \mbox{GeV}^2,
      ~m_{b} = 4.2~   \mbox{GeV},
\end{eqnarray}
where $\Lambda=0.2~ \mbox{GeV}$ is the QCD scale parameter.
In both case (i) and (ii), we plot the parton distribution functions in
the virtual photon with the $\mbox{DIS}_{\gamma}$ scheme~\cite{GRV1992a}.
In addition to the photon PDFs, we evaluate the effective photon structure
function $F^{\gamma}_{\rm eff}(x,Q^2,P^2)$ defined by,
\begin{eqnarray}
 \int_{0}^{1}dx x^{n-2} F^{\gamma}_{2}(x,Q^2,P^2)
  &=& \sum_{i}C^{\gamma}_{i}(n,Q^2)~ q^{\gamma}_{i}(n,Q^2,P^2),\\
 F^{\gamma}_{\rm eff}(x,Q^2,P^2)
 &=&    F^{\gamma}_{2}(x,Q^2,P^2)
            + \frac{3}{2} F^{\gamma}_{L}(x,Q^2,P^2),
\end{eqnarray}
where $C^{\gamma}_{i}(n,Q^2)$'s are the moments of the coefficient
functions in OPE formalism and the index $i$ runs over the related
operators (quarks, gluon, photon), $q^{\gamma}_{i}(n,Q^2,P^2)$'s are the
moments of the photon PDFs, $F^{\gamma}_{2,L}(x,Q^2,P^2)$ are
the usual photon structure functions.
The photon PDFs and the effective photon structure function contain
the massless contribution and the massive contribution (mass effects).
We have evaluated massless contribution up to the NLO in QCD,
the massive contribution is considered up to the LO in QCD,
and we add them together to get the total contributions by using the
formalism adopted in this paper.

\begin{figure}
  \begin{center}
    \def\SCALE{0.45}
    \def\OFFSET{27pt}
    \begin{tabular}{ccc}
      \includegraphics[scale=\SCALE]{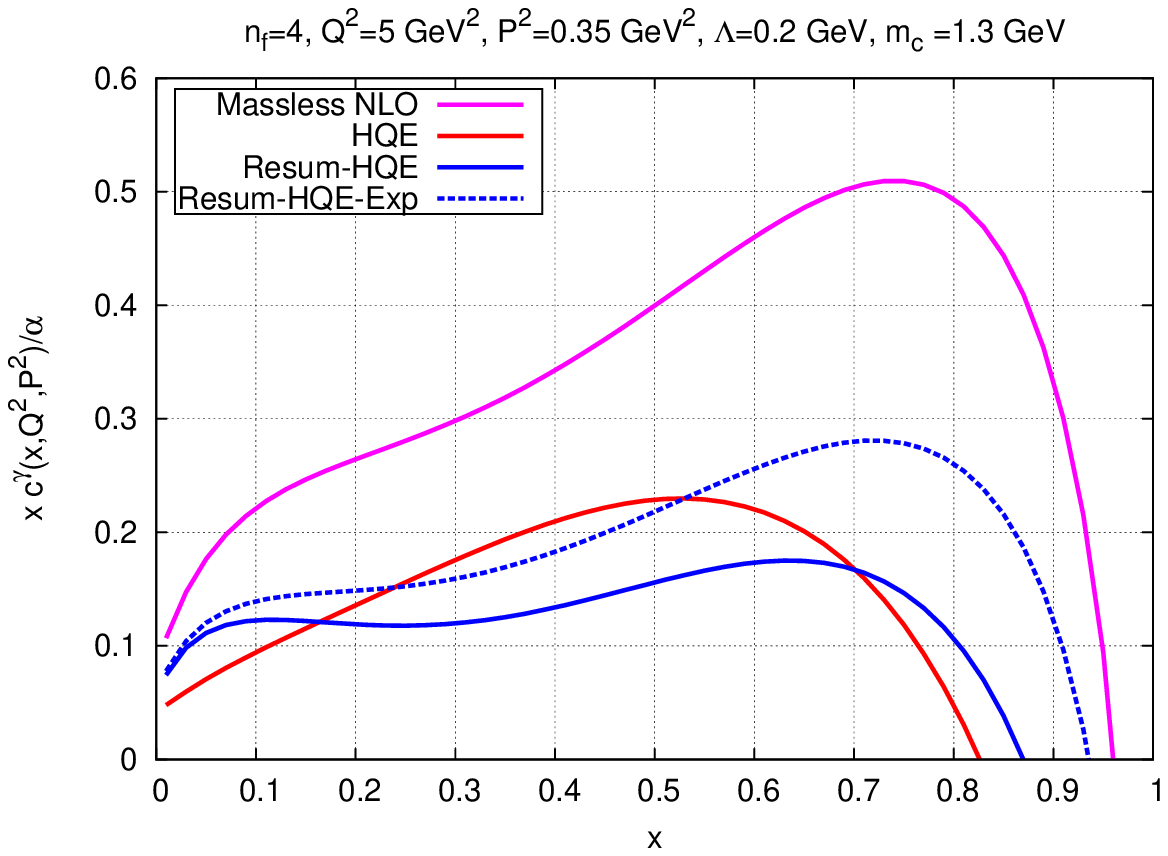} &
      \includegraphics[scale=\SCALE]{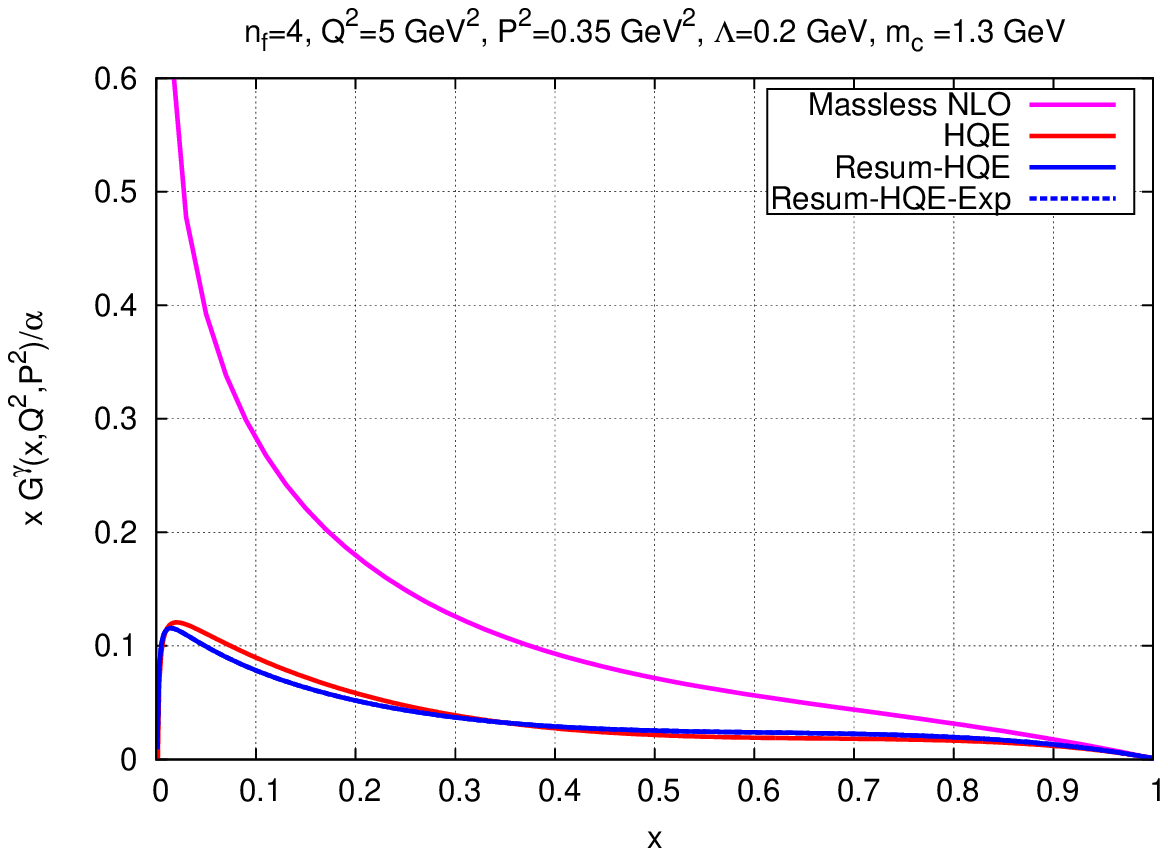} &
      \includegraphics[scale=\SCALE]{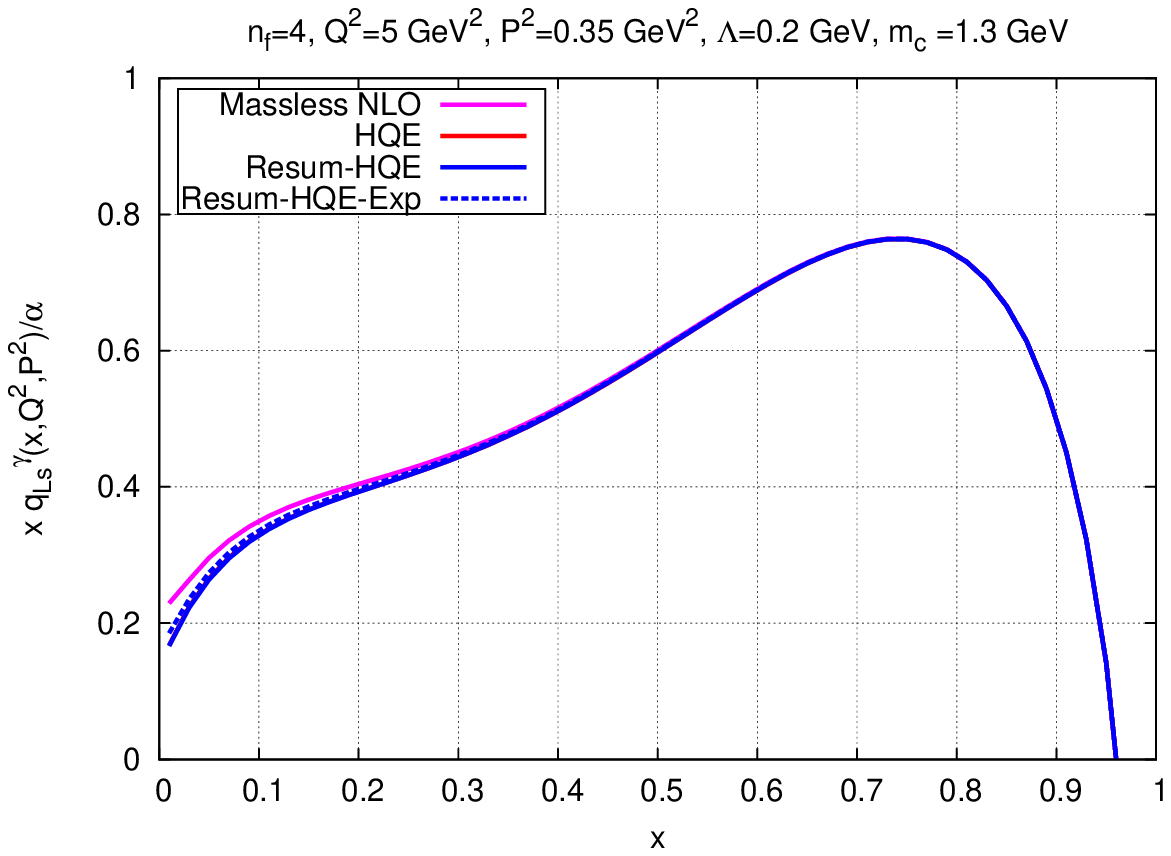} \\
      \hspace{\OFFSET} (a) & \hspace{\OFFSET} (b) & (c) \hspace{\OFFSET}\\
    \end{tabular}
    \caption{%
      Parton distributions in the photon in ${\rm DIS}_\gamma$ scheme
      for $n_f=4$, $Q^2=5~$GeV$^2$, $P^2=0.35~$GeV$^2$ with $m_c=1.3~$GeV
      and $\Lambda=0.2~$GeV:
      (a) $x c^\gamma(x,Q^2,P^2)|_{{\rm DIS}_\gamma}$;
      (b) $x G^\gamma(x,Q^2,P^2)_{{\rm DIS}_\gamma}$,
      and the light singlet quark distribution (c)
          $x q^{\gamma}_{Ls}(x,Q^2,P^2)_{{\rm DIS}_\gamma}.$
    }
    \label{fig_DISg_Q2=5}
  \end{center}
\end{figure}

\begin{figure}
  \begin{center}
    \def\SCALE{0.45}
    \def\OFFSET{27pt}
    \begin{tabular}{ccc}
      \includegraphics[scale=\SCALE]{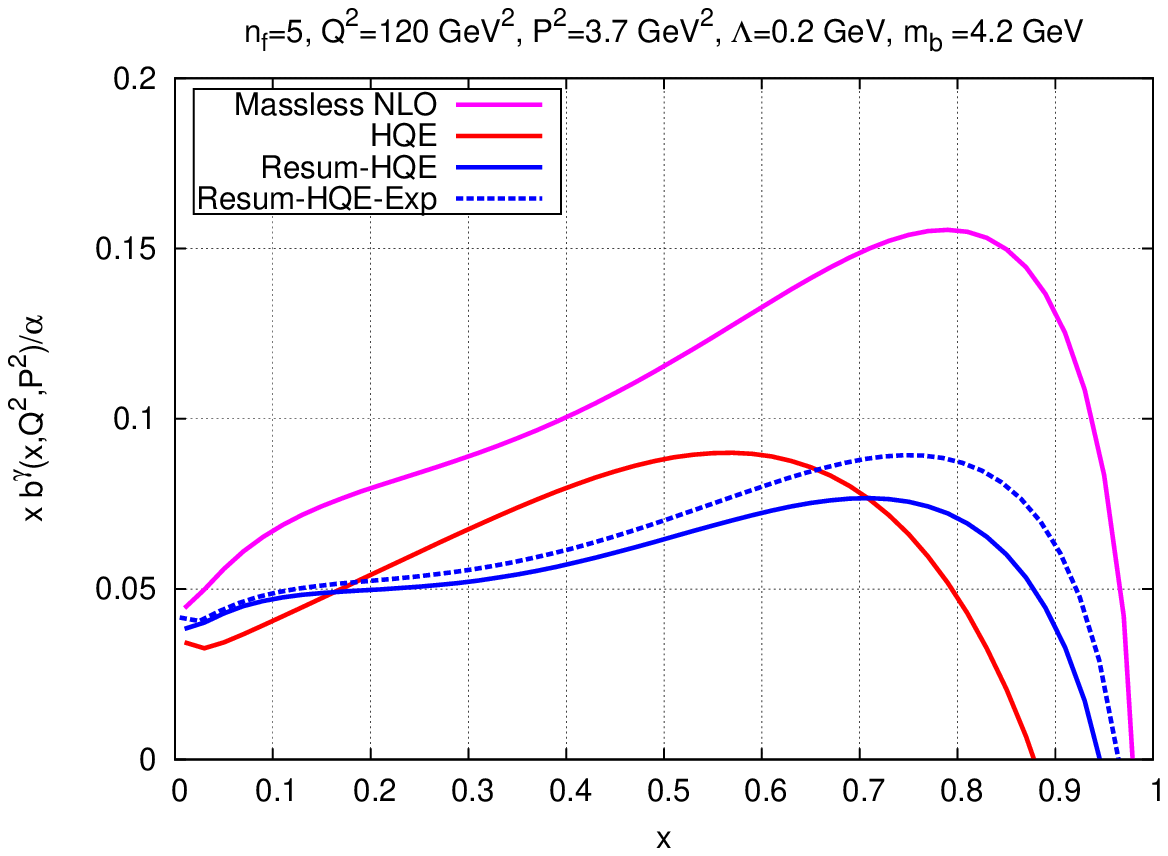} &
      \includegraphics[scale=\SCALE]{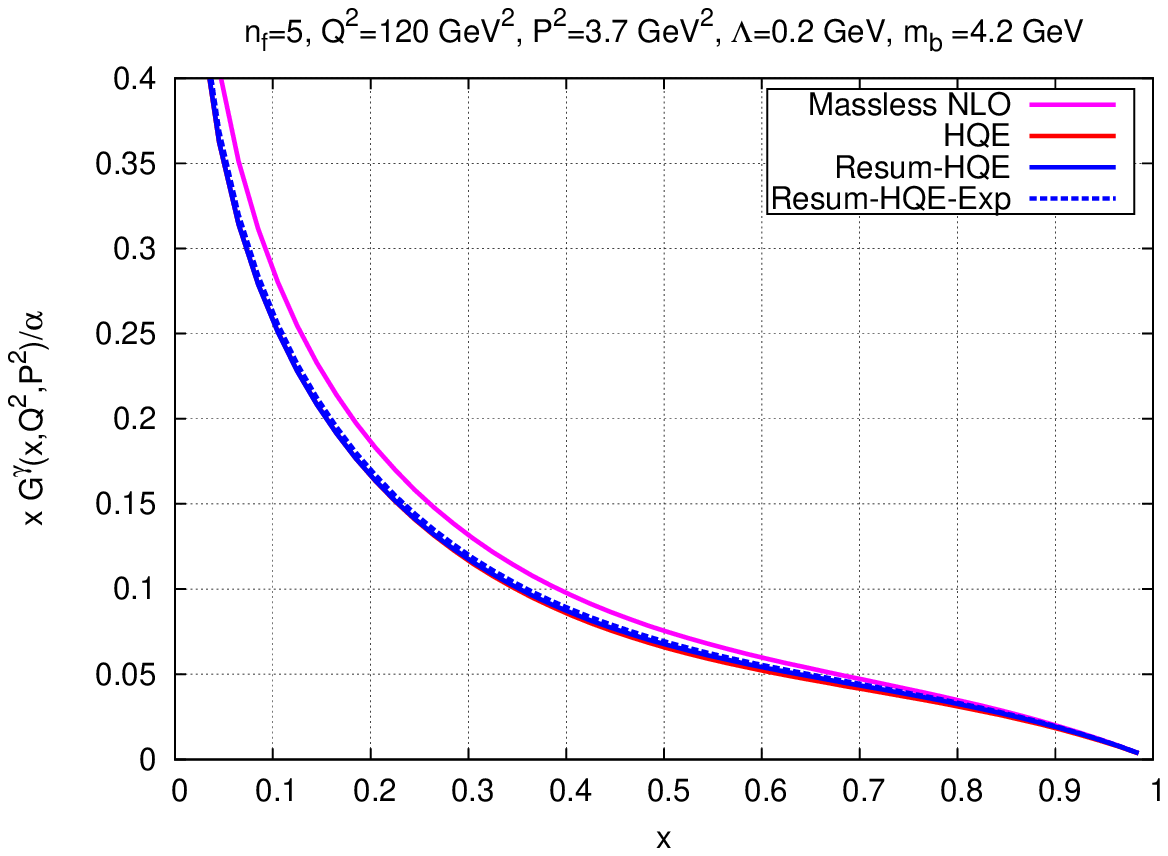} &
      \includegraphics[scale=\SCALE]{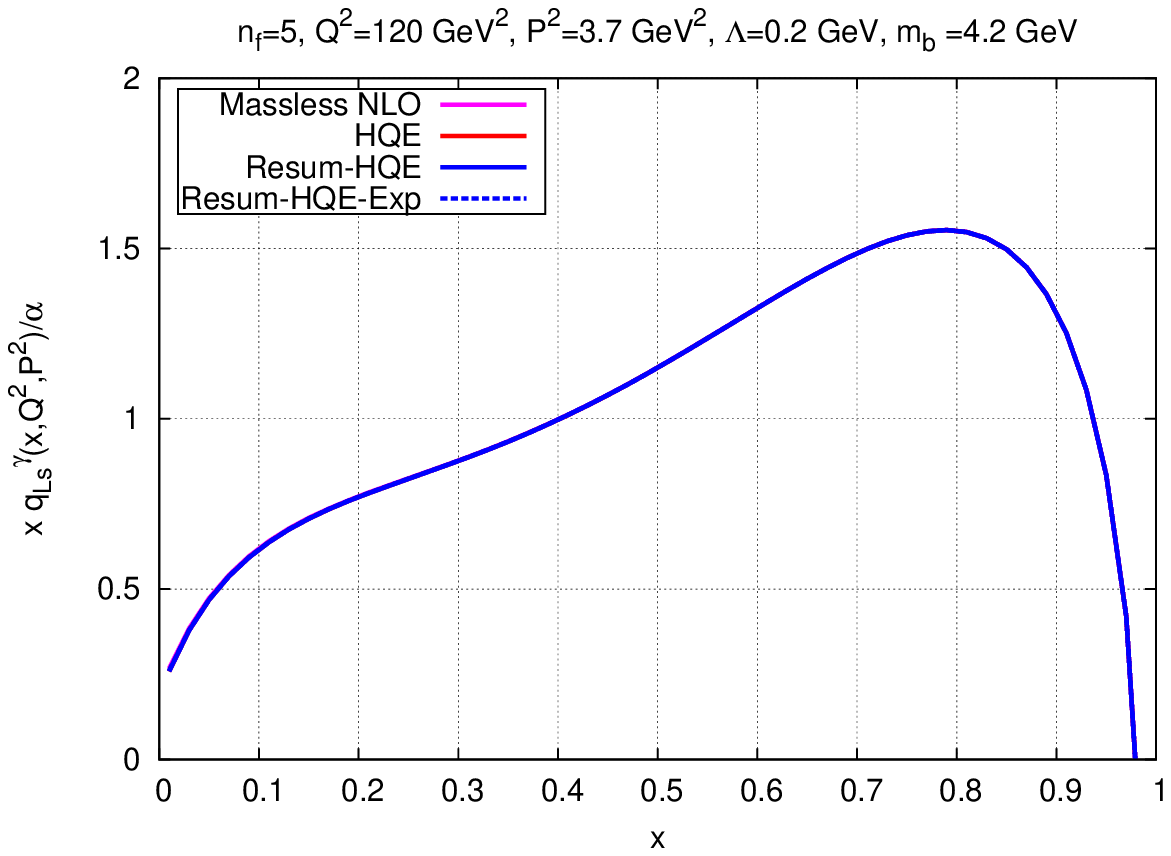} \\
      \hspace{\OFFSET} (a) & \hspace{\OFFSET} (b) & \hspace{\OFFSET} (c)\\
    \end{tabular}
    \caption{%
      Parton distributions in the photon in ${\rm DIS}_\gamma$ scheme
      for $n_f=5$, $Q^2=120~$GeV$^2$, $P^2=3.7~$GeV$^2$ with $m_b=4.2~$GeV
      and $\Lambda=0.2~$GeV:
      (a) $x b^\gamma(x,Q^2,P^2)|_{{\rm DIS}_\gamma}$;
      (b) $x G^\gamma(x,Q^2,P^2)_{{\rm DIS}_\gamma}$,
          and the light singlet quark distribution (c)
          $x q^{\gamma}_{Ls}(x,Q^2,P^2)_{{\rm DIS}_\gamma}.$
    }
    \label{fig_DISg_Q2=120}
  \end{center}
\end{figure}

\begin{figure}[hbt]
  \begin{center}
    \def\SCALE{0.45}
    \def\OFFSET{27pt}
    \begin{tabular}{cc}
      \includegraphics[scale=\SCALE]{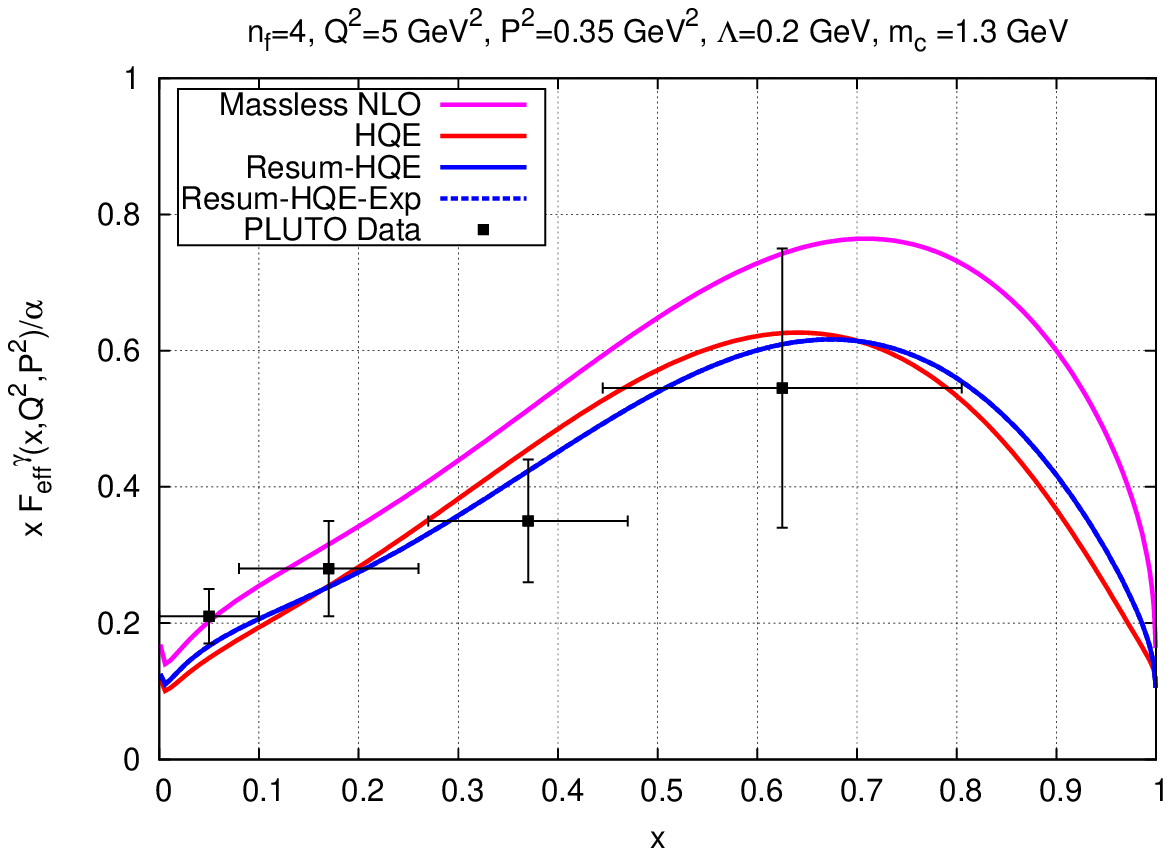} &
      \includegraphics[scale=\SCALE]{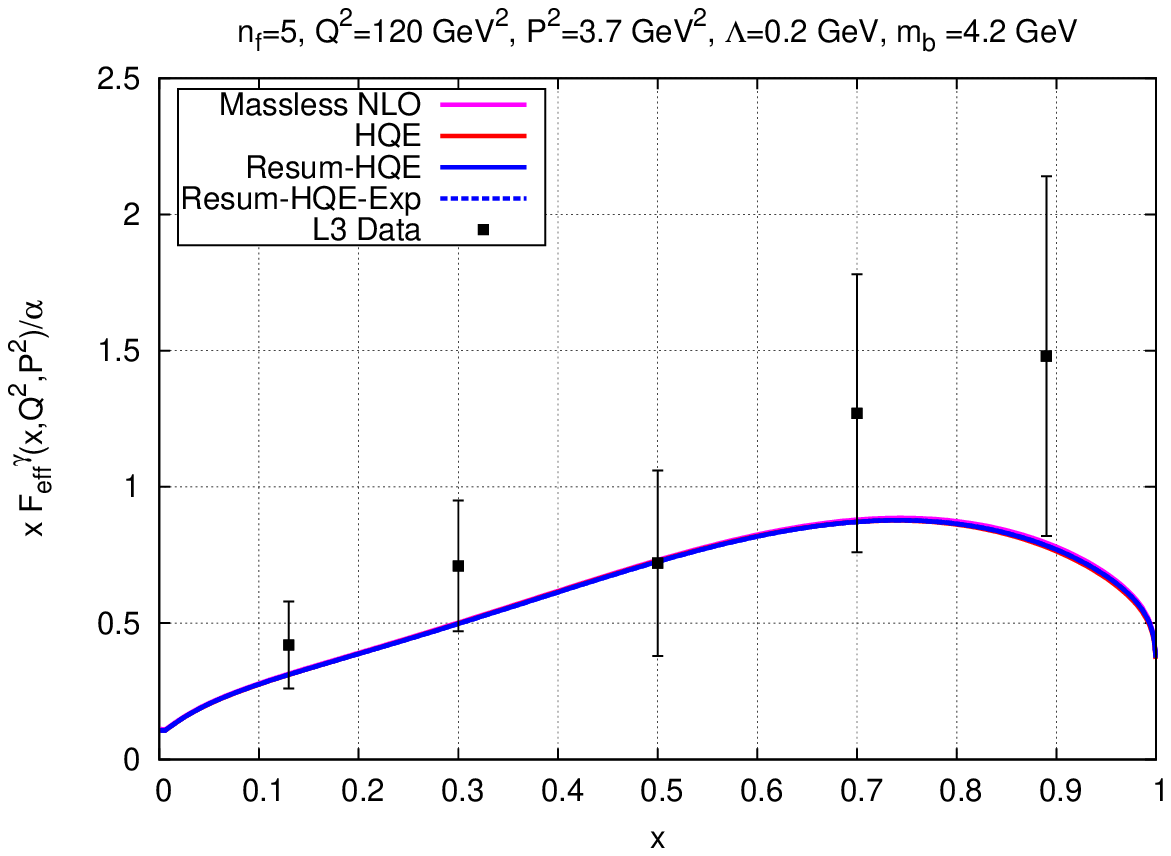} \\
      \hspace{\OFFSET} (a) & \hspace{\OFFSET} (b) \\
    \end{tabular}
    \caption{%
      Effective structure functions
      for 
      PLUTO (a); $n_f=4$, $Q^2=5~$GeV$^2$, $P^2=0.35~$GeV$^2$ 
                 with $m_c=1.3~$GeV and $\Lambda=0.2~$GeV:
      L3    (b) $n_f=5$, $Q^2=120~$GeV$^2$, $P^2=3.7~$GeV$^2$ 
                with $m_b=4.2~$GeV and $\Lambda=0.2~$GeV.
    }
    \label{fig_Feff}
  \end{center}
\end{figure}

We plot (a) the heavy quark PDF $q_{H}^{\gamma}(x,Q^2,P^2)$,
(b) the gluon PDF $G^{\gamma}(x,Q^2,P^2)$, and
(c) the light singlet quark PDF $q^{\gamma}_{Ls}$
in Fig.~\ref{fig_DISg_Q2=5} for the case (i),
and the same three functions in Fig.~\ref{fig_DISg_Q2=120} for the case (ii).
The effective photon structure functions $F^{\gamma}_{\rm eff}(x,Q^2,P^2)$
are plotted in Fig.~\ref{fig_Feff} (a) and (b) for the case (i) and (ii),
respectively.
The abbreviation for the various predictions in the figures are as follows;
`Massless NLO' is the result without the heavy quark effect by using the 
massless OPE formalism,
`HQE' is the result with the heavy quark effect by using the usual OPE
formalism, the `Resum-HQE' is the result with the fully resummed heavy quark
effects, and it is evaluated by the method in which the massless contribution
at the NLO and the heavy quark effects at the LO are combined, and
`Resum-HQE-Exp' is the result with Eq.~(\ref{qh0expand}), respectively.
The difference between `Resum-HQE-Exp' and `HQE' arises from
the mass-independent finite term which one cannot reproduce only
through the renormalisation group equation as we mentioned before.

In general, we can see the large suppression due to the heavy quark mass
effects and its resummation effect on the charm quark distribution (a),
the gluon distribution (b) in Fig.~\ref{fig_DISg_Q2=5}
and the effective photon structure function (a)
in Fig.~\ref{fig_Feff}.
On the other hand, we find a small suppression effect on the bottom quark
distribution (a), the gluon distribution (b) in 
Fig.~\ref{fig_DISg_Q2=120}  and the effective photon
structure function (b) in Fig.~\ref{fig_Feff}.
For the light singlet quark PDF, the heavy quark effect given
in Eq.~(\ref{delta-qls}) is almost negligible, namely the three curves
with heavy quark effects
in Figs.~\ref{fig_DISg_Q2=5}(c) and \ref{fig_DISg_Q2=120}(c)
overlap with each other and they also coincide with the plot \lq
Massless NLO\rq except for the small $x$ region in the case of 
Fig.~\ref{fig_DISg_Q2=5}(c).
This is because the right-hand side of Eq.~(\ref{delta-qls})
can be written as
$(1-1/n_f)2\beta_0\Delta\tilde{A}_{H}^{n}f_n(r)$, where
\begin{eqnarray}
f_n(r)\equiv
-(r)^{d^n_\psi}+\frac{\gamma^{0,n}_{\psi\psi}-\lambda^n_{-}}
{\lambda^n_{+}-\lambda^n_{-}}(r)^{d^n_{+}}+
\frac{\gamma^{0,n}_{\psi\psi}-\lambda^n_{+}}
{\lambda^n_{-}-\lambda^n_{+}}(r)^{d^n_{-}} ,
\end{eqnarray}
which is extremely small as discussed in \cite{KSUU2010}. While
the right-hand side of Eq.~(\ref{delta-qh})
is written as $(1-1/n_f)2\beta_0\Delta\tilde{A}_{H}^{n}g_n(r)$, where
\begin{eqnarray}
g_n(r)\equiv
(r)^{d^n_\psi}+\frac{1}{n_f-1}
\left\{\frac{\gamma^{0,n}_{\psi\psi}-\lambda^n_{-}}
{\lambda^n_{+}-\lambda^n_{-}}(r)^{d^n_{+}}+
\frac{\gamma^{0,n}_{\psi\psi}-\lambda^n_{+}}
{\lambda^n_{-}-\lambda^n_{+}}(r)^{d^n_{-}}\right\}
=(r)^{d^n_\psi}/(1-1/n_f)+f_n(r)/(n_f-1) ,
\end{eqnarray}
which is expressed approximately as $(r)^{d^n_\psi}/(1-1/n_f)$,
and hence the heavy quark mass effects become sizable for heavy
quark PDFs. Note that there exist no heavy quark effects on
light nonsinglet quark PDF $q^{\gamma}_{Lns}$ as seen from
Eq.~(\ref{delta-qlns}).
Since each light quark PDF, $q_L^i$, is a linear combination
of the singlet and nonsinglet quark PDFs, the heavy quark effects
on the light quark PDFs are in fact negligibly small.
Phenomenologically interesting features are the differences
of the size for the heavy quark effect on the heavy quark PDF
and the gluon PDF.
The origin of this difference for the heavy quark effects comes from
their electromagnetic charges. Since the absolute value of the charge of
the charm quark is larger than that of the bottom quark, we obtain the
larger reduction of the gluon PDF in the case (i) than that of in the
case (ii). We can also see these differences in the physical observable;
the effective structure function in Figs.~\ref{fig_Feff} (a)-(b)
in both case (i) and (ii).

The resummation effect of the large logarithmic terms due to
the heavy quark mass is also larger in the case (i) for the three functions
(charm, gluon, effective structure function) than those in the case (ii).
In addition to this feature, we observe a little bit interesting property
of the (a) in Fig.~\ref{fig_Feff}.
It seems to be that the curve with fully resummation effect (Resum-HQE)
slightly close to the experimental data than the curve without the
resummation effect (HQE, Resum-HQE-Exp). It might be due to the validity
of our present method for the resummation of the heavy quark mass effects.
However we cannot say anything about the validity for L3 case
(the figure (b) in Fig.~\ref{fig_Feff}) due to the small
bottom's mass effects on the effective structure function.

Thus, we conclude that the large suppression effect exists in
the theoretical prediction for the charm PDF, gluon PDF in the (b)
with $\mbox{DIS}_{\gamma}$ scheme, the effective photon structure function (a)
of Fig.~\ref{fig_Feff} at the PLUTO's kinematical point not because of
the NLO QCD corrections, but because of the heavy quark (charm quark) effects.
On the other hand, we can see stable results of the theoretical
prediction for the bottom PDF, the gluon PDF in the (b) with
$\mbox{DIS}_{\gamma}$ scheme, the effective photon structure function (b)
of Fig.~\ref{fig_Feff} at the L3's kinematical point
despite the NLO QCD corrections and heavy quark (bottom quark) effects
within this formalism.

\newpage
\section{Conclusion}
\label{conclusion}
We have discussed the resummation of the heavy quark mass effects on the photon
PDFs by changing the initial condition for the LO DGLAP equation. Our
method is based on the mass-independent DGLAP evolution equation
and a change of the initial condition for the heavy quark
component in the LO solution of the DGLAP evolution equation.

We recovered the previous results based on the NLO OPE formalism
\cite{KSUU2010}
except for finite terms by taking a certain limit:
$\ln(m^2/P^2) \ll \ln(m^2/ \Lambda^2$).
By using the method, we can resum the large logarithmic terms due
to heavy quark mass effects on photon PDFs in the virtual photon.
We evaluate the size of the resummation effects numerically.
The resummation effects for the charm quark PDF, gluon PDF are larger
than that of the bottom quark PDF due to the heavy quark electromagnetic
charge. We also see that the resummation of the heavy quark mass effects
on the effective structure function tends to close to the experimental data
in PLUTO case.
Although we only presented the PDFs in the $\overline{\rm DIS}_\gamma$
scheme, the PDFs in the $\overline{\rm MS}$ scheme 
with our resummed heavy quark effects
could be inferred from
the difference between the PDFs in $\overline{\rm MS}$
and those in DIS$_\gamma$ in our previous paper \cite{KSUU2010} where 
both schemes were explicitly presented.

Now some comments on the future extensions of the present work are in order.
One of the possible extensions is the NLO analysis. We can solve the NLO
DGLAP equation including the change of the initial condition for the
LO DGLAP equation. Other possibility
is the extension of this idea
to the $n_f$ quarks system with
two or more heavy quarks.
In the case of two heavy flavours,
this can be achieved by decomposing the $n_f$
quarks into the $n_f-2$ light quarks and two heavy quarks. Such an extension
will be useful to analyse the virtual photon structure functions under the
situation which contains both the massive bottom quark and the charm quark
at Super KEK-B~\cite{KEKB}.
Another example for the phenomenological application of this idea is
to analyse the system with the heavy superpartners in
the supersymmetric QCD.

\begin{acknowledgments}
We would like to thank K.~Sasaki for useful discussions about the mass effect.
This work is supported in part by Grant-in-Aid for Scientific Research
(C) from the Japan Society for the Promotion of Science No.22540276.
\end{acknowledgments}

\appendix
\section{Summary of explicit expressions for the LO solution}
\label{explicit}
We can obtain the expression with the heavy quark effects
for $Ls, H, G, Lns$ components.
Each elements of the LO solution is defined by
\begin{eqnarray}
 \qv_{n}^{\gamma(0)}
= \left( q_{Ls}^{(0),n}(t),~ q_{H}^{(0),n}(t),~ %
         G^{(0),n}(t),~ q_{Lns}^{(0),n}(t)
  \right).
\end{eqnarray}
Then the LO solutions are summarised as
\begin{eqnarray}
 &{}& q_{Ls}^{(0),n}(t) / \frac{\alpha}{8\pi\beta_0} \nn\\
&=& \frac{4\pi}{\alpha_s(t)}
    \left[   \frac{1}{1 + d_{\psi}^{n}}
             \left( \frac{1}{n_f} K_{\psi}^{n} - K_{H}^{0,n}
             \right)
	     \left( 1 - r^{d_{\psi}^{n} + 1} \right)
	   + \frac{n_f-1}{n_f} K_{\psi}^{0,n}
	     \sum_{\pm}\frac{1}{1 + d_{\pm}^{n}}
	     \frac{\gamma_{\psi\psi}^{0,n} - \lambda_{\mp}^{n}}
	          {\lambda_{\pm}^{n} - \lambda_{\mp}^{n}}
	     \left( 1 - r^{d_{\pm}^{n} + 1} \right)
    \right]\nn\\
&{}&  \hspace{0.7cm}+
 \left[  - \frac{n_f-1}{n_f} r^{d_{\psi}^{n}}
         + \frac{n_f-1}{n_f}
           \sum_{\pm}\frac{\gamma_{\psi\psi}^{0,n} - \lambda_{\mp}^{n}}
                {\lambda_{\pm}^{n} - \lambda_{\mp}^{n}} r^{d_{\pm}^{n}}
 \right]  \hat{q}_{H}^{(0),n}(0) / \frac{\alpha}{8\pi\beta_0},\label{Ls}
\end{eqnarray}
\begin{eqnarray}
 &{}& q_{H}^{(0),n}(t) / \frac{\alpha}{8\pi\beta_0} \nn\\
&=& \frac{4\pi}{\alpha_s(t)}
    \left[ - \frac{1}{1 + d_{\psi}^{n}}
             \left( \frac{1}{n_f} K_{\psi}^{n} - K_{H}^{0,n}
             \right)
	     \left( 1 - r^{d_{\psi}^{n} + 1} \right)
	   + \frac{1}{n_f} K_{\psi}^{0,n}
	     \sum_{\pm}\frac{1}{1 + d_{\pm}^{n}}
	     \frac{\gamma_{\psi\psi}^{0,n} - \lambda_{\mp}^{n}}
	          {\lambda_{\pm}^{n} - \lambda_{\mp}^{n}}
	     \left( 1 - r^{d_{\pm}^{n} + 1} \right)
    \right]\nn\\
&{}& \hspace{0.7cm} +
 \left[  \frac{n_f-1}{n_f} r^{d_{\psi}^{n}}
     + \frac{1}{n_f}
       \sum_{\pm}\frac{\gamma_{\psi\psi}^{0,n} - \lambda_{\mp}^{n}}
            {\lambda_{\pm}^{n} - \lambda_{\mp}^{n}}
       r^{d_{\pm}^{n}}
 \right]  \hat{q}_{H}^{(0),n}(0) / \frac{\alpha}{8\pi\beta_0}, \label{H}
\end{eqnarray}
\begin{eqnarray}
  G^{(0),n}(t) / \frac{\alpha}{8\pi\beta_0}
&=& \frac{4\pi}{\alpha_s(t)}
    \left[  K_{\psi}^{n}
            \sum_{\pm} \frac{1}{1 + d_{\pm}^{n}}
	     \frac{\gamma_{G\psi}^{0,n} }
	          {\lambda_{\pm}^{n} - \lambda_{\mp}^{n}}
	     \left( 1 - r^{d_{\pm}^{n} + 1} \right)
    \right]
+ \sum_{\pm}\frac{\gamma_{G\psi}^{0,n} }
             {\lambda_{\pm}^{n} - \lambda_{\mp}^{n}}
         r^{d_{\pm}^{n}} \hat{q}_{H}^{(0),n}(0) / \frac{\alpha}{8\pi\beta_0},
\label{G}
\end{eqnarray}
\begin{eqnarray}
  q_{Lns}^{(0),n}(t) / \frac{\alpha}{8\pi\beta_0}
= \frac{4\pi}{\alpha_s(t)}
     \frac{1}{1 + d_{\psi}^{n}} K_{Lns}^{n}
	     \left( 1 - r^{d_{\psi}^{n} + 1} \right). \label{Lns}
\end{eqnarray}
The first term of the above results corresponds to the massless result
and the second term corresponds to the variation term due the resummed
heavy quark mass effects. Note that there is no extra contribution due
to the modification of the initial condition for the $Lns$ component.
This is consistent with the result in our previous work~\cite{KSUU2010}.
These results are the new and main results in this paper.

\appendix


\end{document}